\documentclass[%
 reprint,
 showpacs,preprintnumbers,
 amsmath,amssymb,
 aps,
 prb,
]{revtex4-1}
\usepackage{dcolumn}
\usepackage[dvipdfm]{graphicx}
\usepackage{subfigure,color}
\usepackage{mathrsfs}

\makeatletter
\def\btt#1{\texttt{\@backslashchar#1}}
\DeclareRobustCommand\bblash{\btt{\@backslashchar}} \makeatother

\begin{document}

\title{Intersurface interaction via phonon in three-dimensional topological insulator}

\author{Tetsuro Habe}
\affiliation{
Department of Applied Physics,
Hokkaido University, Sapporo 060-8628, Japan}

\date{\today}

\begin{abstract}
We theoretically study the phonon mediated intersurface electron-electron interactions on the pseudo two-dimensional metallic states at the two surfaces of a three-dimensional topological insulator.
From a model of a three-dimensional topological insulator including the phonon excitation in it,
we derive the effective Lagrangian which describes the two surface metallic states and the interaction between them.
The intersurface electron-electron interactions can be either repulsive or attractive depending on parameters such as temperature, the speed of the phonon, and the Fermi velocity of the surface states.
The attractive interaction removes the Dirac nodes from the two surface states as a result of the spontaneous symmetry breaking.
On the basis of the calculated results, we also discuss how to tune the inter-surface interaction.
\end{abstract}


\maketitle

\section{introduction}
The three-dimensional(3D) topological insulators(TIs) host two-dimensional(2D) massless electric states on their surface\cite{Fu2007,Fu2007-2,Moore2007,Chen2009}. 
The massless surface state represented by a Dirac Hamiltonian in high-energy physics
 is well preserved as far as the $\mathbb{Z}_2$ topological number being nontrivial in the bulk states in the presence of time-reversal symmetry\cite{Tanaka2009,Hsieh2009,Liu2009,Hor2010,Chen2010,Wray2010,Habe2012}.
The statement, however, is not valid when the electron-electron interactions derive the spontaneous symmetry breaking(SSB).
In fact, a general theory of the Dirac particle\cite{Nambu1961,Nambu1961-2} indicates that 
the mass term appears to the Dirac Hamiltonian in association with the SSB of the chiral $U(1)$ symmetry.
Two key features should be satisfied for the SSB: the presence of two degenerated Dirac cones and the attractive interaction between them. 
According to the theory, the inter-surface interactions are necessary for opening the gap on the surface metallic states of the time-reversal invariant 3D TIs because only a single Dirac cone stays at each surface.

So far, the SSB has been theoretically discussed in a thin film of TI where the inter-surface interactions are mediated by the repulsive Coulomb interaction\cite{Seradjeh2009,Wang2011,LiuM2011,Cho2011,Moon2012,Sodemann2012,Efimkin2012}.
In this situation, to realize the attractive interactions between the two Dirac cones, the large enough applied gate voltage, i.e., the difference of Fermi energies, between the two surfaces is necessary so that electrons appear in one surface and holes appear in the other.
As a result, the attractive interactions between an electron and a hole enable forming the excitonic condensation.
In the real TIs, however, the electron-electron interactions are mainly mediated by phonons
because the dielectric constant in TIs is usually much larger than that in the vacuum.
The effects of the electron-electron interactions via phonon have been a hot issue in the recent studies of 3D TIs\cite{Cheng2011,Giraud2011,Pan2012,Zhu2011,Giraud2012,Zhu2012}.
The theoretical results assuming the interaction via phonons show a good agreement with the experimental results\cite{Cheng2011,Hatch2011}.

In this paper, we theoretically discuss the intersurface interactions of TIs on a simple model in which two quasi two-dimensional surface states  interact with each other via phonons excited in the bulk.
The coupling constant of the inter-surface electron-electron interactions is estimated by the perturbation expansion with respect to the electron-phonon interactions.
We find that the inter-surface interaction via phonons can be attractive 
when the phonon speed comes close to the Fermi velocity of the surface state at low temperature.  
The attractive inter-surface interaction implies a possibility of the phase transition 
from the metallic surface state to the insulating state breaking the chiral $U(1)$ gauge symmetry.

This paper is organized as follows. In Sec.~II, we explain our theoretical model.  
In Sec.~III, we derive the coupling constant of the inter-surface electron-electron interaction.
The comparison between our results and the real materials or the experimental situation is discussed in Sec.~IV.
The conclusion is given in Sec.~V.

\section{Theoretical model}
\begin{figure}[htbp]
 \begin{center}
\includegraphics*[width=80mm]{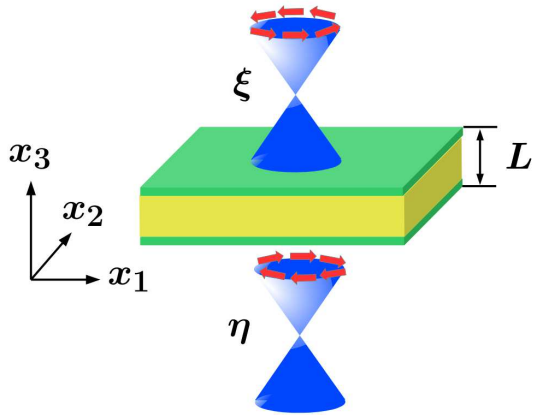}
 \end{center}
\caption{The conceptual scheme of two surface states $\xi$ and $\eta$ which are two-dimensional components at the surface of TIs.
The two cones represent the energy dispersions of the surface states. The arrows on the cones show the direction of spin direction locked by the momentum.
} \label{fig:1}
\end{figure}

We consider the two quasi (2+1)-dimensional surface states  staying on the different surfaces of a 3D TI as shown in Fig.~\ref{fig:1}. 
The three-dimensional wave function is represented by
\begin{align}
\Psi
=&\begin{pmatrix}
f_+(x_3)&0\\
0& f_-(x_3)
\end{pmatrix}\psi\label{3Dwave}
\end{align}
where $f_\pm(x_3)$ localizes at $x_3=\pm L/2$ and exponentially decreases into the bulk insulating region
and 
\begin{align}
\psi=&\begin{pmatrix}
\xi(t,x_1x_2)\\
\eta(t,x_1x_2)
\end{pmatrix},
\end{align} 
contains the spinor wave functions $\xi$ and $\eta$ in a two-dimensional plane as shown in Fig.~\ref{fig:1}.
We assume that the penetration length $\lambda$ is a constant value for each quasi two-dimensional state on the opposite surface.
In this case, the $f_\pm(x_3)$ can be written by,
\begin{align}
f_\pm(x_3)=&\frac{1}{\sqrt{\lambda}}\exp\left[~\mp\frac{1}{2\lambda}\left(x_3\pm \frac{L}{2}\right)~\right].\nonumber
\end{align}
with the thickness $L$ along the $x_3$ axis.
The Lagrangian of the surface states is equivalent to that of the Weyl field as
\begin{align}
\mathscr{L}_{e}=&i\bar{\Psi}\gamma^{\mu}\tilde{\partial}_{\mu}\Psi\label{Weyl},\\
\bar{\Psi}=&\Psi^\dagger \gamma^0\nonumber
\end{align} 
where the index which appears twice in a single term means the summation for $\mu=0,1,2$, and $\gamma^\mu$ is Dirac gamma matrices for $\mu=0-3$, and $5$. 
The $\tilde{\partial}_\mu$ represents the derivative for time $\partial_0=\tilde{\partial}_0=\partial/\partial t$ and in-plane coordinates $\tilde{\partial}_j=v_F\partial_j=v_F\partial/\partial x_j$ for $j=1$ and $2$ with the Fermi velocity $v_F$.
The Lagrangian has no mass term represented by $m\gamma^3$ or $m\gamma^5$ with a constant $m$.
In the presence of time-reversal symmetry, the mass term proportional to $\gamma^3$ vanishes because it represents magnetization or magnetic field.
The other mass term proportional to $\gamma^5$ mixes the two surface states $f_+$ and $f_-$ on either surface in Eq.~(\ref{3Dwave}).
In the absence of electron-electron interactions, the mass term is negligible in a thick TI because $\lambda/L\ll1$ is satisfied.
Namely, the direct hopping between the two two-dimensional metallic states on the opposite surface is suppressed.
Therefore the inter-surface interaction is intermediated only by the phonon field with a large dielectric constant.

\subsection{Effective 2D Lagrangian of electron-phonon coupling}
We estimate the coupling constant of the interaction between the pseudo two-dimensional electric states and the three-dimensional phonon.
In the bulk of the TI, the electron-phonon interaction between electron and phonon can be ignored 
because the Fermi energy lies in the insulating gap.
Thus, in the bulk of TIs, the three-dimensional phonon can be written by the bare scalar boson field $\Phi$.
The scalar boson's Lagrangian is a Klein-Gordon-type one,
\begin{align}
\mathscr{L}_{p}=\left(\partial_{0}\Phi\right)^2-\sum_{j=1-3}\left(v_B\partial_{j}\Phi\right)^2-\frac{m^2}{2}\Phi^2
,\label{Lp}
\end{align}
with the speed $v_B$ and the mass $m$ of the scalar boson.
The mass term (the third term in Eq. (\ref{Lp})) appears in the case of the optical phonon or the acoustical phonon under pressure\cite{Cheng2011}.
The bare scalar boson field $\Phi$ has the eigen energy $\omega_k=\sqrt{{v_B}^2|\boldsymbol{k}|^2+m^2}$ with the momentum $\boldsymbol{k}$.
The wave function is represented by a product of the plane wave along the $x_3$ axis and the in-plane wave component of $\phi(t,x_1,x_2,k_3)$ as
\begin{align}
\Phi(t,x)=\int \frac{dk_3}{\sqrt{2\pi}}\phi(t,x_1,x_2,k_3)\frac{1}{\sqrt{2\pi}}\sin k_3x_3,
\end{align}
under the Dirichlet boundary condition of $\Phi(t,x)=0$ at $x_3=0$.

At the vicinity of the surface, the amplitude of the electron-phonon coupling is much larger than that in the bulk 
because of the metallic property at the surface.
The Lagrangian of the electron-phonon interaction is calculated by 
the overlap integration of the wave functions for the surface states and that of the phonon,
\begin{align}
L_{ep}=\int d^3\boldsymbol{x}U&\left\{\xi^\dagger(x)\xi(x){f_{+}}^2(x_3)\right.\nonumber\\
&\left.+\eta^\dagger(x)\eta(x){f_{-}}^2(x_3)\right\}\Phi,\label{Lep}
\end{align}
where $U$ is the coupling function and the overlap is integrated in $(x_1,x_2,x_3)$.
In the limit of a small penetration length $\lambda$ as $L/\lambda\rightarrow\infty$ and $1+k_3\lambda\rightarrow1$,
the overlap integration along the $x_3$-axis can be written by the correction of the coupling constant $\chi$:
\begin{align}
\chi=&\frac{1}{2\pi\lambda}\int^{L}_{0}dx_3\sin k_3x_3\exp\left[-\frac{x_3}{\lambda}\right]\nonumber\\
=&\frac{1}{2\pi} \frac{1}{1+(k_3\lambda)^2}\left(e^{-\frac{L}{\lambda}}[\sin k_3L-k_3\lambda\cos k_3L]+k_3\lambda\right)\nonumber\\
\simeq&\frac{k_3\lambda}{2\pi}.
\end{align}
Thus, the interaction Lagrangian $L_{ep}$ is independent of $L$,
\begin{align}
L_{ep}=&\int dx_1dx_2~\tilde{\phi}(x)\{\xi^\dagger(x)\xi(x)+\eta^\dagger(x)\eta(x)\},\label{efLep}\\
\tilde{\phi}(x)=&\int dk_3\chi(k_3) U(x,k_3)\phi(x,k_3),
\end{align}
with $x=(t,x_1,x_2)$.

\section{Intersurface interaction of Weyl field}
From the interaction Lagrangian in Eq. (\ref{efLep}), 
we will derive the effective intersurface interaction
represented by
\begin{align}
V=-g~\xi^\dagger(x)\xi(x)~\eta^\dagger(x)\eta(x),\label{int}
\end{align}
where the coupling constant $g$ reflects the dynamics of the phonon and depends on temperature, phonon mass, and the ratio of the Fermi velocity to the phonon speed.

\begin{figure}[htbp]
 \begin{center}
\includegraphics*[width=80mm]{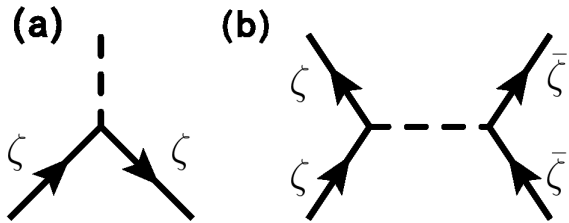}
 \end{center}
\caption{The Feynman diagrams of (a) the vertex function in Eq.(\ref{Lep}) and (b) the lowest order term for electron-electron interaction via the phonon exchanging are shown.
The label $\zeta$ denotes the surface Weyl field for $\zeta=\xi,~\eta$.
In Fig. (b), $\zeta$ and $\bar{\zeta}$ represent opposite surface field operators, {i.e.}, $(\zeta,\bar{\zeta})=(\xi,\eta)$ or $(\zeta,\bar{\zeta})=(\eta,\xi)$.
The diagram of (b) represents the first-order effective inter-surface interaction through a phonon.
} \label{fig:SE1}
\end{figure}

In the lowest order of the perturbation expansions with respect to the phonon field,
the bare inter-surface interaction is represented by the Lagrangian of
\begin{align}
\mathscr{L}_{\mathrm{int}}=-\int dk_3&{u}^2(x,k_3)\xi^\dagger(x)\xi(x)\nonumber\\
&\times D(x-x')\zeta^\dagger(x')\zeta(x'),\label{Lint}
\end{align} 
with the bare vertex function $u^2(x,k_3)=\chi^2U(x,k_3)U(x',k_3)$ and the phonon propagator $D(x-x')$ (see Appendix~\ref{AP-effective action}).
The Lagrangian of interaction can be interpreted easily as the exchange of phonons between the two electrons staying on the different surfaces.
The event represented by the Lagrangian is schematically shown by using a Feynman diagram in Fig.\;\ref{fig:SE1}-(b).

\begin{figure}[htbp]
 \begin{center}
\includegraphics*[width=80mm]{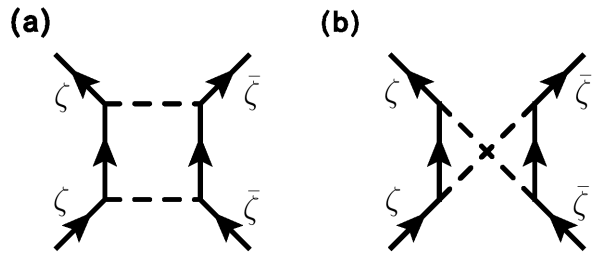}
 \end{center}
\caption{The Feynman diagrams of the second order electron-phonon interaction are shown.
} \label{fig:SE2}
\end{figure}

The effective inter-surface interaction in Eq. (\ref{int}) can be calculated by a summation of the bare intersurface interaction in Eq.~(\ref{Lint}) and the one-particle irreducible(1PI) vertex function $\kappa^{[\mathrm{1PI}]}$
\begin{align}
V=-\mathscr{L}_{\mathrm{int}}-\kappa^{[\mathrm{1PI}]},
\end{align}
where the 1PI vertex function contains the loop diagrams consisting of the bare phonon propagators and fermion propagators e.g. the diagrams in Fig. \ref{fig:SE2}.
To calculate the coupling constant in Eq.~(\ref{int}),
we consider the effective potential instead of the one-loop 1PI vertex function as shown in Fig.~\ref{fig:SE2}.
The effective potential is the vertex function under the condition 
that the initial state and the final state have the same energy and the same momentum.
In this case, the interaction between any states $\xi$ and $\eta$
in momentum space
can be represented by using a single coupling constant $g$ as
\begin{align}
V=-g(p')\xi^\dagger(p_1)\xi(p_1+p')\eta^\dagger(p_2)\eta(p_2-p')\label{eq:15}
\end{align}
where $p=(p_0,\boldsymbol{p})$ is the vector with the energy $p_0$ and the two-dimensional momentum $\boldsymbol{p}=(p_1,p_2)$.
The intersurface interaction with $p'=0$ is dominant in the low energy theory,
even if the coupling constant contains no divergent term in the infrared region. 
At $p'=0$, the Fourier transformation of Eq.~(\ref{eq:15}) is essentially equivalent to Eq.~(\ref{int}).
Therefore, the coupling constant $g$ in Eq.~(\ref{int}) is calculated from $g(0)$ in Eq.~(\ref{eq:15}). 

We introduce the propagators of the Weyl field, those of the phonon, and the bare vertex function $u=\chi U$ in momentum space.
The propagators of the Weyl field $G(p_0,\boldsymbol{p})$ and the phonon $D(k_0,\boldsymbol{k})$
 in energy-momentum space are represented by
\begin{align}
G(p_0,\boldsymbol{p})=&\frac{1}{p_0\sigma^0-v\boldsymbol{p}\cdot\boldsymbol{\sigma}+i\epsilon},\\
D(p_0,\boldsymbol{p})=&\frac{2\sqrt{{v_B}^2\boldsymbol{p}^2+\tilde{m}^2}}{{p_0}^2-{v_B}^2\boldsymbol{p}^2-(\tilde{m}^2-i\epsilon)}\label{phonon},\\
\tilde{m}=&m+{v_B}^2{k_3}^2
\end{align}
where $\sigma^{\nu}$ is $2\times2$ unit matrix for $\nu=0$ and
Pauli matrix for $\nu=1-3$.
We consider two types of the electron-phonon coupling: the deformation coupling and polar one\cite{Mahan2000}.
The piezoelectric coupling is suppressed by inversion symmetry of the materials\cite{Huang2008,Giraud2011}.
Thus we ignore the piezoelectric coupling in this paper. 
According to Ref. \onlinecite{Mahan2000}, the two electron-phonon couplings are characterized by the different bare vertex functions $M_p^{(D)}$ and $M_p^{(P)}$ for the deformation and polar couplings, respectively.
The bare vertex functions are the Fourier components of $u$ and are represented by
\begin{align}
\left(M_p^{(D)}\right)^2=&C_1\chi^2\frac{v_F|\boldsymbol{p}|^2}{p_D}\label{CPd}\\
\left(M_p^{(P)}\right)^2=&C_2\chi^2\sqrt{{v_B}^2|\boldsymbol{p}|^2+\tilde{m}^2}\label{CPp},
\end{align}
where $C_1$ and $C_2$ are the coupling constants.
The coupling constants are
\begin{align}
C_1=&\frac{p_D}{v_F}R^2\\
C_2=&2\pi e^2\left(\frac{1}{\varepsilon_{\infty}}-\frac{1}{\varepsilon}\right)
\end{align}
where $R$ is the deformation constant, $p_D$ is a Debye momentum, and $\varepsilon_{\infty}$ and $\varepsilon$ are the dielectric constants for a high-frequency and a low-frequency, respectively\cite{Mahan2000}.
We consider both couplings in this paper because it is unclear which coupling is more dominant in TIs.
The deformation coupling comes from the interaction between the electron and the acoustical phonon. On the other hand, the polar coupling accounts for the interaction between the electron and the optical phonon.
Some experiments suggest that the optical phonon play a dominant role\cite{Shahil2010,Qi-J2010}, 
however another experiment\cite{Hatch2011} on the life-time of a quasi-particle has good agreement with the theory of the acoustical phonon\cite{Giraud2011}.

\subsection{Deformation coupling}

\begin{figure}[htbp]
 \begin{center}
\includegraphics*[width=80mm]{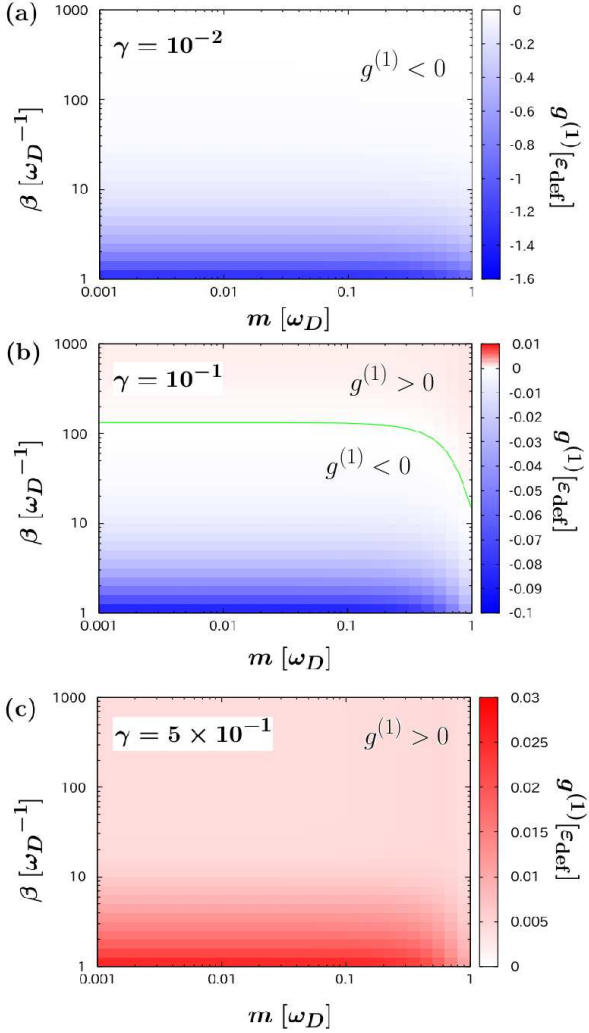}
 \end{center}
\caption{The coupling constant $g^{(1)}$
of the inter-surface interaction via deformation in Eq.~\ref{effp} is shown as a function of temperature $\beta^{-1}$ and the phonon mass $m$.
The potential becomes attractive in the red region 
and repulsive in the blue region.
The zero-potential is represented by the green line.
} \label{fig:4}
\end{figure}

At first, we calculate the coupling constant $g$ for the interaction via the deformation coupling. 
In the no-loop calculation, the deformation coupling constant vanishes, 
\begin{align}
\mathscr{L}_{\mathrm{int}}=&g^{(0)}\xi^\dagger\xi(p)\zeta^\dagger\zeta(-p)\\
g^{(0)}=&-\lim_{p_\mu\rightarrow0}\int dk_3{M_p^{(D)}}^2D(p_0,\boldsymbol{p})=0,
\end{align}
where the superscript of $g^{(0)}$ denotes the number of loops.
Thus we consider the one-loop effective potentials $V_a^{(1)}$ and $V_b^{(1)}$, where the subscript denotes two diagrams in Fig.~\ref{fig:SE2}.
The one-loop effective potentials are represented by
\begin{align}
V_a^{(1)}=&\int dk_3\int \frac{d^3p}{(\sqrt{2\pi})^3}{M_p^{(D)}}^4\left(\xi^\dagger G(p_0,\boldsymbol{p})\xi\right)\nonumber\\
&\times\left(\eta^\dagger G(-p_0,-\boldsymbol{p})\eta\right)D(p_0,\boldsymbol{p})^2\\
V_b^{(1)}=&\int dk_3\int \frac{d^3p}{(\sqrt{2\pi})^3}{M_p^{(D)}}^4\left(\xi^\dagger G(p_0,\boldsymbol{p})\xi\right)\nonumber\\
&\times\left(\eta^\dagger G(p_0,\boldsymbol{p})\eta\right)D(p_0,\boldsymbol{p})^2,
\end{align}
where $\xi$ and $\eta$ are independent of $(p_0,\boldsymbol{p})$.
The net effective potential is
\begin{align}
V^{(1)}=\int dk_3\int \frac{d^3p}{(\sqrt{2\pi})^3}&{M_p^{(D)}}^4(~
{p_0}^2\Gamma_1(p)\xi^\dagger\xi~\eta^\dagger\eta\nonumber\\
&+{p_i}^2\Gamma_2(p)\xi^\dagger\sigma^i\xi~\eta^\dagger\sigma^i\eta~
).
\label{inp}
\end{align}
Using the Matsubara method, only the spin independent term remains at finite temperature $T=\beta^{-1}$,
\begin{align}
V^{(1)}=-g^{(1)}\xi^\dagger\xi~\eta^\dagger\eta\label{effp},
\end{align}
where the second term in Eq.~(\ref{inp}) disappears.
The effective coupling constant $g^{(1)}$ is
\begin{align}
g^{(1)}=&\frac{1}{2}\int dk_3\int dp~p(\omega_p(M_p^{(D)})^2)^2\Gamma_1(p)\label{eq27}\\
\Gamma_1(p)=&\Gamma_{FB}(p)+\Gamma_F(p)+\Gamma_B(p)\nonumber\\
\Gamma_{FB}(p)=&\beta[\tilde{D}_A(p)^2+\tilde{D}_R(p)^2]\nonumber\\
&\times[n_F(\varepsilon_p)n_F(-\varepsilon_p)+n_B(\omega_p)n_B(-\omega_p)]\\
\Gamma_F(p)=&-\frac{1}{\varepsilon_p}\left([1-4{\varepsilon_p}^2\tilde{D}_A(p)]\tilde{D}_A(p)^2n_F(\varepsilon_p)\right.\nonumber\\
&\left.-[1-4{\varepsilon_p}^2\tilde{D}_R(p)]\tilde{D}_R(p)^2n_F(-\varepsilon_p)\right)\\
\Gamma_B(p)=&\frac{1}{\omega_p}\left([1+4{\omega_p}^2\tilde{D}_A(p)]\tilde{D}_A(p)^2n_B(\omega_p)\right.\nonumber\\
&\left.-[1+4{\omega_p}^2\tilde{D}_R(p)]\tilde{D}_R(p)^2n_B(-\omega_p)\right)
\end{align}
where $\omega_p=\sqrt{{v_B}^2|\boldsymbol{p}|^2+\tilde{m}^2}$ and $\varepsilon_p=v_F|\boldsymbol{p}|$ are the energy of the phonon and the surface Weyl field respectively. 
The integral range of momentum is $0<\sqrt{p^2+{k_3}^2}<p_D$ with the Debye momentum $p_D$.
The Bose and Fermi distribution functions $n_B(z)$ and $n_F(z)$ are represented by
\begin{align*}
n_B(z)=&\frac{1}{e^{\beta z}-1},\\
n_F(z)=&\frac{1}{e^{\beta z}+1}.
\end{align*}
The propagator $\tilde{D}_{A(R)}(p)$ is defined by
\begin{align}
\tilde{D}_{A(R)}(p)=\frac{1}{{\omega_p}^2-{\varepsilon_p}^2-(+)i\epsilon}.
\end{align}
In Fig.~\ref{fig:4}, we show $g^{(1)}$ calculated numerically with a unit of $\varepsilon_{\mathrm{def}}={C_1}^2p_D/v_F$ as a function of three parameters $\beta$, $\gamma=v_B/v_F$, and $m$; $\beta$ and $m$ are normalized by using the energy $\omega_D=\sqrt{m^2+{v_B}^2{p_D}^2}$.
In the real TIs, the $\gamma$ is approximately estimated as $10^{-2}$; i.e., the phonon speed is much smaller than the surface Fermi velocity $v_B\ll v_F$.
For small $\gamma$, the phonon contributes to the repulsive intersurface interaction as shown in Fig.~\ref{fig:4}-(a).
Increasing the speed ratio $\gamma$, however, the region of the repulsive intersurface interaction is decreasing in Fig.~\ref{fig:4}-(b).
Eventually, the intersurface interaction changes to an attractive one
under any mass $m$ and temperature $\beta^{-1}$ in Fig.~\ref{fig:4}-(c).

\subsection{Polar coupling}

\begin{figure}[htbp]
 \begin{center}
\includegraphics*[width=80mm]{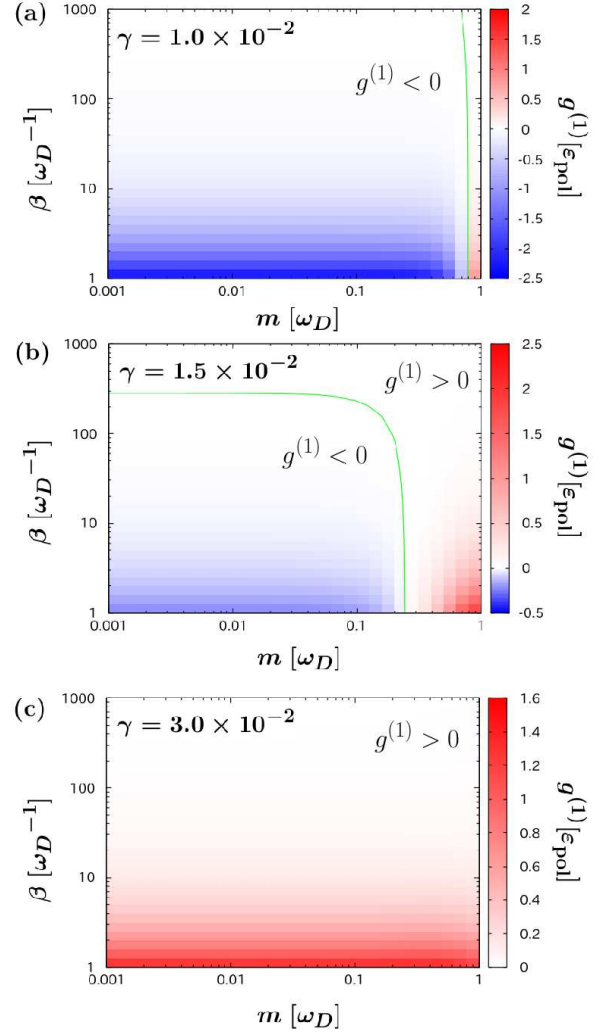}
 \end{center}
\caption{The coupling constant $g^{(1)}$ with an unit of $\varepsilon_{\mathrm{pol}}={C_2}^2p_D/v_F$ of the inter-surface interaction via the polar coupling in Eq.~\ref{effp} is shown.
The potential is attractive in red region and repulsive in the blue region.
The zero-potential is represented by the green line.
} \label{fig:5}
\end{figure}

Next, we consider the coupling constant $g$ of the interaction via the polar coupling.
The interaction is attractive in the no-loop approximation, 
\begin{align}
g^{(0)}=&\lim_{p_\mu\rightarrow0}\int dk_3{M_p^{(P)}}^2D(p_0,\boldsymbol{p})=2{C_2}\kappa,\label{opt}\\
\kappa=&\frac{\lambda^2}{8{v_B}^3}\left(
v_Bp_D{\omega_D}(2{\omega_D}^2-m^2)-m^4\ln\frac{v_Bp_D+\omega_D}{m}
\right),\nonumber
\end{align}
where $\kappa$ is always a positive value.
The no-loop term of the phonon has the opposite sign to that of the photon, i.e., the electromagnetic field(see Appendix~\ref{AP1}).
We also calculate the one-loop correction $g^{(1)}$ with a unit of $\varepsilon_{\mathrm{pol}}={C_2}^2p_D/v_F$ 
shown in Fig.~\ref{fig:5}.
For the polar coupling, the crossover between a repulsive interaction and an attractive one occurs at lower $\gamma$ than that of the deformation coupling.

\begin{figure}[htbp]
 \begin{center}
\includegraphics*[width=80mm]{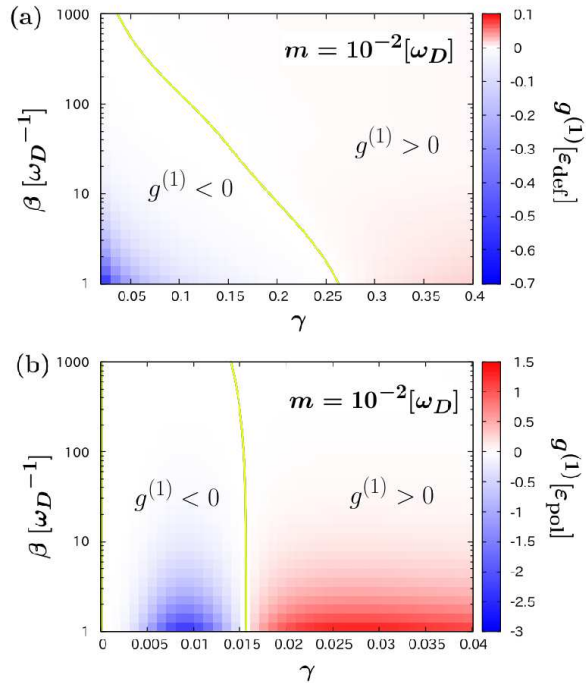}
 \end{center}
\caption{The coupling constants of the inter-surface interaction via the two couplings are shown.
These figures are calculated by use of (a) deformation and (b) polar coupling with $m=10^{-2}[\omega_D]$.
The interaction becomes attractive in the red region and repulsive in the blue region, and the zero-potential is represented by the green line.
} \label{fig:6}
\end{figure}

It is difficult to decide which interaction via deformation or polar coupling is more dominant 
because the coupling constants $C_i$ in Eqs.~(\ref{CPd}) and (\ref{CPp}) are unknown in TIs.
According to our calculation, the coupling constants in both cases become attractive 
when the materials have the slow Dirac mode or the fast phonon mode i.e. the speed ratio $\gamma$ is larger than the ordinary 3D TIs of Bi$_2$Se$_3$, Bi$_2$Te$_3$, and Sb$_2$Te$_3$.
In Fig.~\ref{fig:6}, the coupling constant $g^{(1)}$ is shown as a function of $\gamma$ and the temperature $T=\beta^{-1}$.
The coupling constant $g^{(1)}$ must become attractive when the speed ratio $\gamma$ is larger than $0.25$ for the deformation coupling and $0.015$ for the polar coupling.

At the end of this section, we show the resultant Lagrangian density describing the electric states on the opposite surface,
\begin{align*}
\mathscr{L}=i\bar{\psi}\gamma^{\mu}\tilde{\partial}_{\mu}\psi(x)+\frac{g}{4}\left[\{\bar{\psi}\psi(x)\}^2+\{\bar{\psi}i\gamma^{5}\psi(x)\}^2\right].
\end{align*}
where the coupling constant of the interaction is equal to the summation of $g=g^{(0)}+g^{(1)}$.
The model Lagrangian density is so-called Nambu-Jona Lasino model in 2+1 dimension $x=(t,x_1,x_2)$.
The model shows a dynamical mass generation, i.e. a dynamical gap generation, for a large positive $g$ in association with 
the spontaneous symmetry breaking of chiral $U(1)$ gauge symmetry\cite{Nambu1961,Nambu1961-2}.

\section{Discussion}
We show that the intersurface interaction via phonon opens a gap on the energy dispersion of the surface electric states in association with the spontaneous symmetry breaking.
The dynamical gap generation occurs when the inter-surface interaction is attractive.
In this section, we compare the conditions for the attractive interaction in the present theory with those of real materials.
The Fermi velocity of the surface states on 3D TIs, e.g., Bi$_2$Se$_3$, Bi$_2$Te$_3$ and Sb$_2$Te$_3$ is approximately estimated as $v_F$ $10^{4}$-$10^{5}$[m/s] in the first principles calculation\cite{Liu2010}.
The phonon speed $v_B$ in 3D TIs is approximately estimated to be $10^{3}$[m/s].
Therefore, since $\gamma=v_F/v_B$ is $10^{-2}$-$10^{-3}$, 
the attractive intersurface interaction is possible in such materials when the polar coupling is dominant.
Even when the deformation coupling is dominant, the attractive coupling is possible at low temperature below sub-Kelvin.
Generally speaking, the Fermi velocity is proportional to the strength of a spin-orbit interaction.
Therefore, the dynamical gap generation may also occur in TIs whose spin-orbit coupling is weak enough to have large $\gamma$.

\section{Conclusion}
We have studied the intersurface electron-electron interaction mediated by phonon propagating in the bulk region of a three-dimensional topological insulator.
The sign and magnitude of the coupling constant for the interaction strongly depend on parameters such as temperature, the phonon mass, and the ratio of the Fermi velocity of surface Dirac states to the phonon speed. 
When the phonon speed is much smaller than the Fermi velocity,
the intersurface interaction is repulsive.
When the phonon speed is not so much smaller than the Fermi velocity, on the other hand,  
the intersurface interaction can be attractive within the accessible temperature.
The attractive intersurface interaction opens a gap on the metallic surface states 
in association with the appearance of the spontaneous symmetry breaking electric states at the surface.

\begin{acknowledgments}
I would like to thank Yasuhiro Asano for helpful comments. 
\end{acknowledgments}

\appendix

\section{Effective action of Weyl field}\label{AP-effective action}
In this appendix, we derive the effective intersurface interaction like Eq.~(\ref{Lint}) from the Lagrangian of Eqs.~(\ref{Lp}) and (\ref{efLep}) by the path-integral method.
The effective action which contains the effective interaction is acquired from the generating functional of a propagator $Z$
 represented by
\begin{align}
Z[z,\bar{z},J]&=\int \mathscr{D}\psi\mathscr{D}\bar{\psi}\mathscr{D}\phi
\exp i\int d^3x \mathscr{L}\\
\mathscr{L}&=
\mathscr{L}_e+\mathscr{L}_p+u_0(\bar{\psi}\gamma_0\psi)\phi
+J\phi+\bar{z}\psi+z\bar{\psi},\label{Lagrangian}
\end{align}
where $z(x)$, $\bar{z}(x)$ and $J(x)$ are the virtual external sources and we introduce the coupling constant $u_0$.
The effective action is the generating functional of the vertex function\cite{Negele1998}.
To omit an explicit phonon field, we consider that there is no phonon source as $J=0$ and calculate the path-integral for $\phi$,
\begin{align}
Z[z,\bar{z}]=&\int \mathscr{D}\psi\mathscr{D}\bar{\psi}
\exp\left[ i\int d^3x \mathscr{L}_e'\right]Z_0[j(x)],\\
\mathscr{L}_e'=&\mathscr{L}_e+\bar{z}\psi+z\bar{\psi},\nonumber\\
Z_0[j(x)]=&\exp\left[-i\int d^3x\int  d^3x'j(x)
D(x-x')j(x')\right],\nonumber\\
j(x)=&u(\bar{\psi}\gamma_0\psi)(x),\nonumber
\end{align}
with the phonon propagator $D(x)$.
The calculation is the same as in the case of a non-interacting scalar boson with external source of $j(x)$ 
and provides the generating functional of the phonon propagator as $Z_0[j(x)]$.
We obtain the four point term in $Z_0$ which represents an effective interaction in the Weyl field,
\begin{align}
\mathscr{L}_{\mathrm{int}}=-j(x)D(x-x')j(x')\label{AP2-1}.
\end{align}
The interaction can be interpreted easily as exchanging the phonon.
The Lagrangian contains an intra-surface interaction proportional to the fourth-order term of $\xi$ or $\eta$ in Eq.~(\ref{int}) 
and an intersurface interaction which consists of the product of the quadratic term of $\xi$ and the quadratic term of $\eta$. 
The effective action can be derived as a summation of one-particle irreducible vertex functions which consist of the four-field potential in Eq.~(\ref{AP2-1}).

\section{The Coulomb interaction}\label{AP1}
The interaction Lagrangian via a photon in Feynman gauge is represented by
\begin{align}
\mathscr{L}_{\mathrm{int}}=&-e^2\bar{\psi}\gamma^{\mu}\psi D_{\mu\nu}(p)\bar{\psi}\gamma^{\nu}\psi\\
D_{\mu\nu}(p)=&\frac{-g_{\mu\nu}}{{p_0}^2-|\boldsymbol{p}|^2+i\varepsilon}
\label{photon},
\end{align}
where the speed of light is unity $c=1$ and $g_{\mu\nu}$ is the metric tensor as
\begin{align}
g_{\mu\nu}=\mathrm{diag}[1,-1,-1,-1],
\end{align}
with $\mu,~\nu=0-3$.
Using the longitudinal component given by $D_{00}(p)$, the interaction without energy transfer $p_0=0$ can be represented by
\begin{align}
\mathscr{L}_{\mathrm{int}}=&-(\xi^\dagger\xi+\eta^\dagger\eta)\frac{e^2}{|\boldsymbol{p}|^2}(\xi^\dagger\xi+\eta^\dagger\eta)\\
=&-\frac{e^2}{|\boldsymbol{p}|^2}
\left(\sum_{\zeta=\xi,\eta}\zeta^\dagger\zeta\zeta^\dagger\zeta
+2\xi^\dagger\xi\eta^\dagger\eta\right).
\end{align}
The first and second terms are the intra-surface and inter-surface Coulomb interactions, respectively.
This interaction is repulsive in contrast to that mediated by phonons in Eq.~(\ref{opt}) because the photon propagator in Eq.~(\ref{photon}) is opposite in sign to the phonon propagator in Eq.~(\ref{phonon}).

\bibliography{TI}

\end{document}